\documentclass[aip,jcp,reprint]{revtex4-1}
\usepackage{graphicx}
\usepackage{amsmath}
\usepackage{amssymb} 
\usepackage{mhchem}

\begin{document}
\title{Yukawa particles in a confining potential}

\author{Matheus Girotto}
\email{matheus.girotto@ufrgs.br}
\affiliation{Instituto de F\'isica, Universidade Federal do Rio Grande do Sul, Caixa Postal 15051, CEP 91501-970, Porto Alegre, RS, Brazil}

\author{Alexandre P. dos Santos}
\email{alexandreps@ufcspa.edu.br}
\affiliation{Departamento de Educa\c c\~ao e Informa\c c\~ao em Sa\'ude, Universidade Federal de Ci\^encias da Sa\'ude de Porto Alegre, 90050-170, Porto Alegre, RS, Brazil.}
\affiliation{Departamento de F\'isica, Universidade Federal de Santa Catarina, 88040-900, Florian\'opolis, Santa Catarina, Brazil}

\author{Thiago Colla}
\email{thiago.colla@ufrgs.br}
\affiliation{Faculty of Physics, University of Vienna, Boltzmanngasse 5, A-1009 Vienna, Austria}

\author{Yan Levin}
\email{levin@if.ufrgs.br}
\affiliation{Instituto de F\'isica, Universidade Federal do Rio Grande do Sul, Caixa Postal 15051, CEP 91501-970, Porto Alegre, RS, Brazil}

\begin{abstract}

We study the density distribution of repulsive Yukawa particles confined by an external potential. In the weak coupling limit, we show that the mean-field theory is able to accurately account for the particle distribution. In the strong coupling limit, the correlations between the particles become important and the mean-field theory fails. For strongly correlated systems, we construct a density functional theory which provides an excellent description of the particle distribution, without any adjustable parameters. 

\end{abstract}

\maketitle

\section{Introduction}

The Yukawa potential is used to model interparticle interactions in plasmas~\cite{ZaFu10,Ba10}, 
dusty plasmas~\cite{ToKi98,FoIv05,DzRa13}, colloidal suspensions~\cite{MeFr91b,HeHo11,vava13}, and atomic physics~\cite{JiHo13,CeWi13}.  
In soft-matter systems, the exponential screening of the effective potential 
arises from the positional  
correlations between the oppositely charged particles~\cite{Li01,Le02}. 
Because of its great importance for various models, the thermodynamics of Yukawa systems has been a subject of extensive study~\cite{HaFa91,HaFr94,ScVa13}. Most of the previous work, however, has been
restricted to the homogeneous fluid or solid states~\cite{RoKr88,MeFr91a,LoPa93,PaMo95,StRo98,HoRo04,GaNa12}.
In this paper we will investigate a gas of Yukawa particles  confined by an external potential. 
Such situation arises, for example,  when a colloidal system is acted on by the electromagnetic
field produced by the laser 
tweezers~\cite{CrGr96,Cr97,Gr97,DuGr98,LiYu00,Lo01,Re03,ElGu10}. 
Without loss of generality, in this paper we will consider the external potential which has a 
one dimensional parabolic form
\begin{equation}\label{w}
W(z)=\dfrac{\alpha z^2}{2} \ ,
\end{equation}
where $\alpha$ is a measure of the trap strength. 
The theory developed below, however, can be
applied to an arbitrary confining potential $W(x,y,z)$.  

We will first show that in the weak-coupling limit (high temperatures) the density distribution 
of Yukawa gas is well described by the mean-field~(MF) theory~\cite{LePa11,GiDo13}. In the strong coupling limit (low temperature), the positional correlations between the particles become important and the MF theory fails~\cite{RoHa97,AlCh84,Le98}.   In this case we will construct a density functional theory~(DFT) based on the \textit{hypernetted-chain}~(HNC) equation and the \textit{local density approximation}~(LDA) and will show that this theory  accounts very accurately for the particle  distribution. All the theoretical results will be compared with the Monte Carlo~(MC) simulations.

\section{Mean-Field Theory}

We study a system of $N$ particles interacting through a repulsive Yukawa potential
\begin{equation}\label{V}
V(r)=q G({\bf r}_1,{\bf r}_2) \ ,
\end{equation} 
where
\begin{equation}\label{green}
G({\bf r}_1,{\bf r}_2)=q \dfrac{e^{-\lambda r}}{r} \ ,
\end{equation}
$r=|{\bf r}_1-{\bf r}_2|$, $\lambda$ is the typical inverse distance, and $q$ is the strength of the interaction potential.   For colloidal systems, $q$ is
\begin{equation}\label{qColl}
q=\frac{Ze}{\sqrt{\epsilon_w}},
\end{equation} 
where $Ze$ is the  charge of colloidal particles, $e$ is the proton charge, $\epsilon_w$ is the dielectric constant of the medium, and  $\lambda$ is the inverse Debye length
which depends on the ionic strength inside the suspension~\cite{Le02}.

We first observe that $G({\bf r}_1,{\bf r}_2)$  satisfies the Helmholtz equation
\begin{equation}\label{gfe}
\nabla^2G({\bf r},{\bf r}_1)-\lambda^2 G({\bf r},{\bf r}_1)=-4\pi q \delta({\bf r}-{\bf r}_1) \ .
\end{equation}
Consider a  Yukawa gas confined to a hyperstripe with periodic boundary conditions in the $x$ and $y$ directions
and open in the $z$ direction. The solution of Eq.~\eqref{gfe} for such system can be expressed as
\begin{equation}\label{gf}
G({\bf r},{\bf r}_1)=\frac{2 \pi q}{L_x L_y}\sum_{m_x,m_y}e^{2\pi i\left[ \dfrac{m_x}{L_x}(x-x_1)+\dfrac{m_y}{L_y}(y-y_1)\right]}\dfrac{e^{-\gamma_m|z-z_1|}}{\gamma_m} \ ,
\end{equation}
where
\begin{equation}
\gamma_m=\sqrt{\lambda^2+4\pi^2(\frac{m_x^2}{L_x^2}+\frac{m_y^2}{L_y^2})} \ ,
\end{equation}
and $-\infty<m_x<\infty$ and $-\infty<m_y<\infty$ are integers. $L_x$ and $L_y$ are the widths of the hyperstripe in the $x$ and $y$ directions, respectively.

In equilibrium, the distribution of confined particles is given by 
\begin{equation}\label{om}
\rho(z)=A  e^{-\beta \omega (z)} \ ,
\end{equation}
where $\beta=1/k_B T$, $\omega(z)$ is the potential of mean force~(PMF), and $A$ is the normalization constant~\cite{Le02}. In the weak-coupling limit, the correlations between the particles can be neglected and the PMF can be approximated by  $\omega(z)=q\phi(z)+W(z)$, where $\phi(z)$ is the Yukawa potential at position $z$. This constitutes a MF approximation for the particle distribution,
\begin{equation}\label{dens}
\rho(z)=A  e^{-\beta[q\phi(z)+W(z)]} \ ,
\end{equation}
where
\begin{equation}
A=\frac{N}{L_x L_y\int_{-\infty}^{+\infty} dz \ e^{-\beta[q\phi(z)+W(z)]}} \ .
\end{equation}
The potential $\phi(z)$ can be expressed in terms of the Green's function, Eq.~\eqref{gf},
\begin{equation}\label{phi1}
\phi( z)=\int d{\bf r}' \rho( z')G({\bf r},{\bf r}') \ .
\end{equation}
Integrating over $x$ and $y$ coordinates, Eq.~\eqref{phi1} simplifies to
\begin{equation}\label{phi2}
\phi(z)=\frac{2\pi q}{\lambda}\int_{-\infty}^{+\infty} dz' \rho(z')e^{-\lambda |z'-z|} \ .
\end{equation}
Substituting Eq.~\eqref{dens} into Eq.~\eqref{phi2}, we obtain an integral equation for the mean potential.   This equation can be solved numerically using Picard iteration. 

To test the accuracy of the MF theory we perform MC simulations. $N=100$ Yukawa particles are confined in a box of sides $L_x$, $L_y$ and $L_z$, with periodic boundary conditions in $x$ and $y$ directions. In the $z$ direction the particles are constrained by an external potential $W(z)$. The periodic lengths are  taken to be $L_x=L_y=35~\lambda^{-1}$, while the cutoff for the particle-particle interaction is set at $10~\lambda^{-1}$. In the Metropolis algorithm, a new  configuration $n$ is constructed from an old configuration $o$ by a small displacement of a random particle. The new state is accepted with a probability $P=\min\{1,e^{-\beta(E_n-E_o)}\}$, where $E_n$ and $E_o$ are the energies of the new and the old configurations, respectively. If the movement is not accepted, the configuration $o$ is preserved and counted as a new state. The length of the displacement is adjusted during the simulation in order to obtain the acceptance rate of $50\%$. The energy of the system used in the MC simulations is given by
\begin{equation}
E=\sum_{i=1}^{N-1}\sum_{j=i+1}^{N}q G({\bf r}_i,{\bf r}_j)+\sum_{i=1}^{N}W(z_i) \ .
\end{equation}
The averages are calculated using $10^5$ uncorrelated states, obtained after $10^6$ MC steps for equilibration. To quantify the strength of the particle-particle interaction  and the trap-particle interaction, 
it is convenient to define the following dimensionless parameters
\begin{equation}
\epsilon=\frac{q^2\lambda}{k_BT} \quad\mbox{and}\quad \chi=\frac{\alpha}{k_BT\lambda^2} \ .
\end{equation} 

\begin{figure}[h]
\vspace{0.2cm}
\includegraphics[width=8.cm]{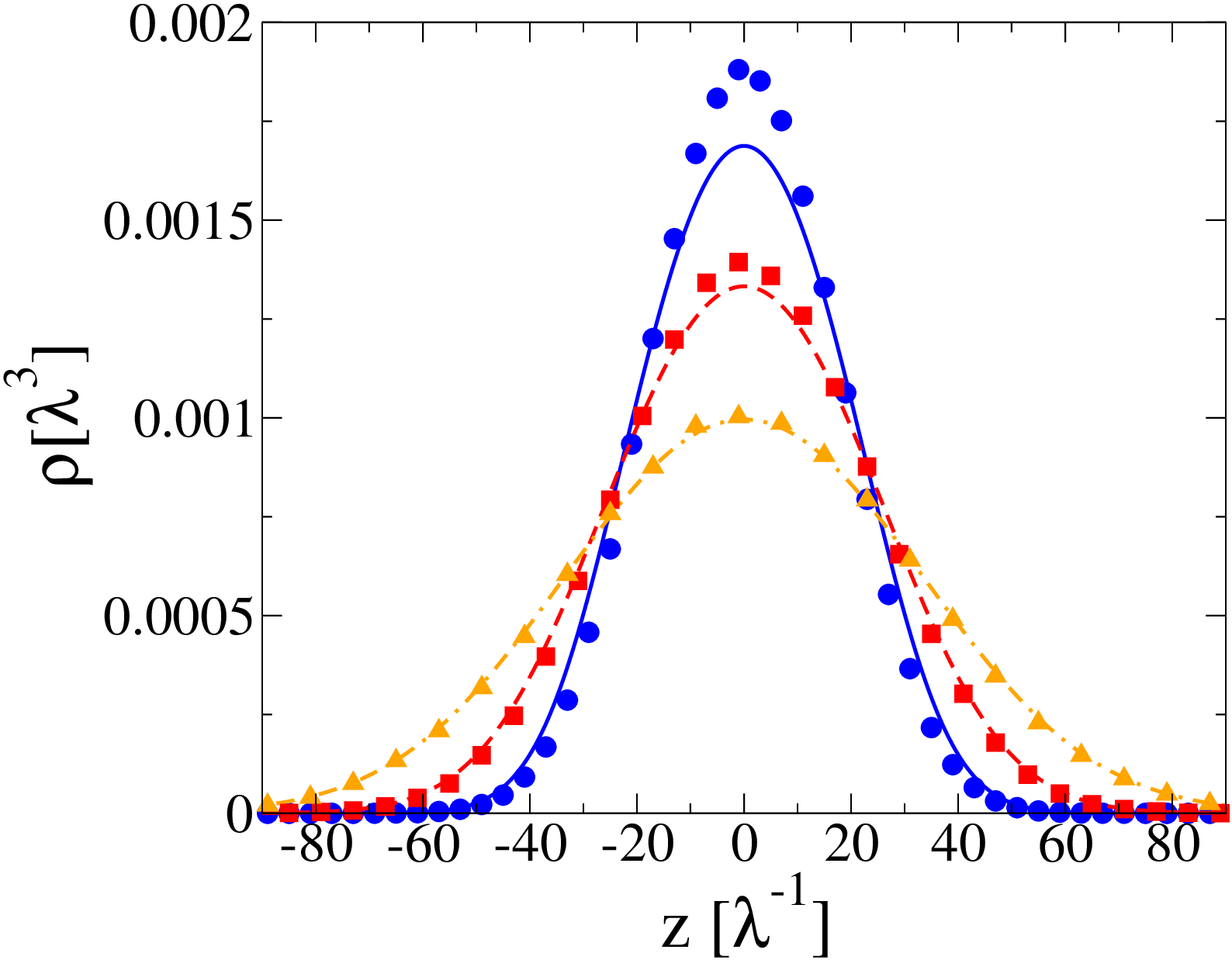}\vspace{0.2cm}
\caption{Density profile of Yukawa gas with $\epsilon=40$ and $\chi=0.004$ (circles); $\epsilon=20$ and $\chi=0.002$ (squares); and  $\epsilon=10$ and $\chi=0.001$ (triangles). 
The symbols represent MC simulation data, while the curves are the predictions of MF theory. }
\label{fig1}
\end{figure}

We can now compare the solutions of the MF equations \eqref{dens} and \eqref{phi2} with the results of MC simulations, see Fig.~\ref{fig1}. For high temperatures --- low values of $\epsilon$  --- the MF theory accounts very well for the particle distribution observed in MC simulations.  On the other hand, in the strong coupling limit (low temperatures), the correlations between the particles become important and the MF theory starts to fail. Positional correlations between the particles lead to greater occupation of the low energy states than is predicted by the MF theory~\cite{Le02}. This is similar to the process of overcharging observed in 
colloidal suspensions with multivalent ions~\cite{Sh02,PiBa05,DoLe10}.

\section{Density functional theory}

The failure of the MF theory to properly account for the density distribution of a confined 
Yukawa gas is a consequence of strong positional correlations between the particles at low temperatures. To account for these correlations, we appeal to the DFT.  The
equilibrium particle distribution corresponds to the minimum of the Helmholtz free energy
\begin{equation}\label{F1}
F[\rho(z)]=F^{ent}+F^{int}+F^{cor}\ 
\end{equation}
subject to the constraint
\begin{equation}\label{F2}
\int d{\bf r} \rho({\bf r})=N \ .
\end{equation}
In Eq.~\eqref{F1}, $F^{ent}$ is the entropic contribution to the free energy, $F^{int}$ is the interaction part (which includes both the MF interaction and the interaction with the external potential), and $F^{cor}$ is the correlational free energy. In general, the correlational free energy is a  non-local function of density $\rho(z)$.  For systems with hard-core interactions this requires development
of sophisticated weighted density approximations~\cite{Ro89,RoSc96,Ev09,Ro10,FrLe13}. For repulsive Yukawa particles, however,
the density variation should be much smoother and we expect that a LDA for $F^{cor}[\rho(z)]$ will be sufficiently accurate. LDA assumes that the system achieves a local thermodynamic equilibrium within a range smaller than the typical length scale of the system inhomogeneity ~\cite{Ev79,HaMc06}. This condition is fulfilled provided that the density distribution does not vary dramatically. Since the density profiles resulting from the soft potential, Eq.~\eqref{w}, are smooth (see Fig.~\ref{fig1}), we expect that the LDA will be sufficiently  accurate in the present situation as long as $\epsilon$ and $\chi$ are not too large.
Performing the minimization of the total free energy, we obtain the equilibrium particle density distribution~\cite{Le02},
\begin{equation}\label{dens2}
\rho(z)=N\frac{e^{-\beta\left[q\phi(z)+W(z)+\mu^{cor}[\rho(z)]\right]}}{L_x L_y\int_{-\infty}^{+\infty} dz \ e^{-\beta\left[q\phi(z)+W(z)+\mu^{cor}[\rho(z)]\right]}} \ ,
\end{equation}
where the correlational chemical potential is
\begin{equation}\label{mu1}
\mu^{cor}[\rho(z)]=\frac{\delta F^{cor}}{\delta \rho(z)} \ .
\end{equation}
Within the LDA, $\mu^{cor}[\rho(z)]$ is calculated using the free energy of a homogeneous system 
\begin{equation}
\mu^{cor}[\rho(z)]=\frac{\partial f^{cor}(\bar{\rho})}{\partial \bar{\rho}}\biggr\arrowvert_{\bar{\rho}=\rho(z)} \ ,
\label{mu_local}
\end{equation}
where $f^{cor}$ is the correlational free energy density of a homogeneous Yukawa system.
When the correlations are negligible  (high temperatures), $\mu^{cor}$ vanishes and the MF theory, Eq.~\eqref{dens}, becomes exact.

To calculate the correlational chemical potential we use the HNC equation. This equation is known to 
account well for the structural and thermodynamic properties of Yukawa-like systems \cite{At96,CoDo12}. The HNC approximation is based on a closure relation
\begin{equation}\label{clo1}
h(\mathbf{r})=\ln\left[-\beta v(\mathbf{r})+h(\mathbf{r})-c(\mathbf{r})\right]-1 ,
\end{equation}
for the Ornstein-Zernike equation, where  
$h(\mathbf{r})$ is the pair correlation function,   $c({\bf r})$  is the direct correlation function,  and $v(\mathbf{r})$ is the particle-particle interaction potential. 
In the Fourier space, the Ornstein-Zernike equation, for an isotropic system, takes a particularly simple form,
\begin{equation}\label{OZ1}
h(k)=\frac{c(k)}{1-\rho c(k)}.
\end{equation}
This equation can be solved iteratively. First, we make an initial guess for the direct correlation function, $c_0(r)$.  The  Fourier transform of $c_0(r)$ is then inserted into Eq.~\eqref{OZ1}, yielding a zero order approximation of $h_0(k)$.  The inverse Fourier transformation provides $h_0(r)$. The closure relation, Eq.~\eqref{clo1}, allows us to calculate the next order direct correlation function, $c_1(r)$, etc. The process is repeated until convergence is achieved~\cite{HaMc06}. To speed up the convergence, a method of Ng  with six parameters is used for updating the $c(r)$ at each iteration step~\cite{Ng74}. 

\begin{figure}[t]
\vspace{0.2cm}
\includegraphics[width=8.cm]{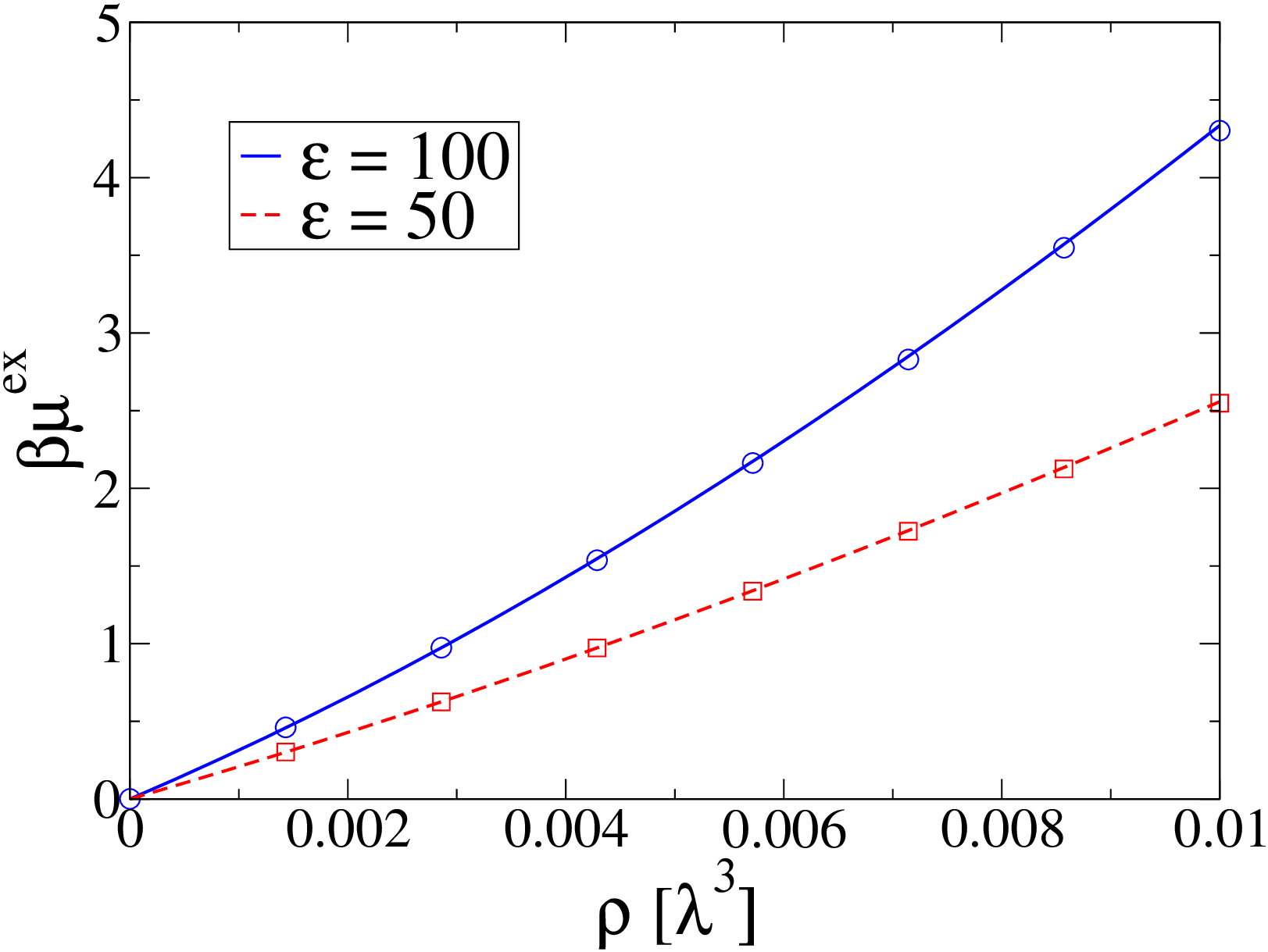}\vspace{0.2cm}
\caption{The excess (over the ideal gas) chemical potential as a function of the particle concentration for $\epsilon=100$ and $\epsilon=50$. The symbols represent the simulation data, and the lines are calculated using the HNC equation and Eq.~\eqref{mu2}.}
\label{fig2}
\end{figure}

Within the HNC approximation, the excess (over the ideal gas) chemical potential~\cite{HaMc06} is 
\begin{equation}\label{mu2}
\beta \mu^{ex}=\frac{1}{2}\rho\int h({\bf r})[h({\bf r})-c({\bf r})]d{\bf r}-\rho\int c({\bf r})d{\bf r} \ .
\end{equation}
In Fig.~\ref{fig2}, we show that the Eq.~\eqref{mu2} agrees perfectly  with the chemical  potential calculated using the MC simulations and Widom particle insertion algorithm~\cite{FrSm01}.

The excess chemical potential contains both the MF  and the correlational contributions. 
The correlational  chemical potential, $\mu^{cor}$, is calculated by 
subtracting from $\mu^{ex}$ the MF  part
\begin{equation}
\mu^{mf}=\dfrac{\partial F^{mf}}{\partial N} \ ,
\end{equation}
where the MF free energy of a homogeneous Yukawa gas is
\begin{equation}\label{fmf}
F^{mf}=\dfrac{ q^2}{2} \rho^2 \int d^3{\bf r} \int d^3{\bf r}'\dfrac{e^{-\lambda|{\bf r}-{\bf r}'|}}{|{\bf r}-{\bf r}'|} \ .
\end{equation}
Integrating Eq.~\eqref{fmf} and then differentiating with respect to $N$, we obtain 
\begin{equation}\label{mu3}
\beta \mu^{mf}=\frac{4\pi\rho q^2 \beta}{\lambda^2} \ .
\end{equation}

\begin{figure}[t]
\includegraphics[width=8.cm]{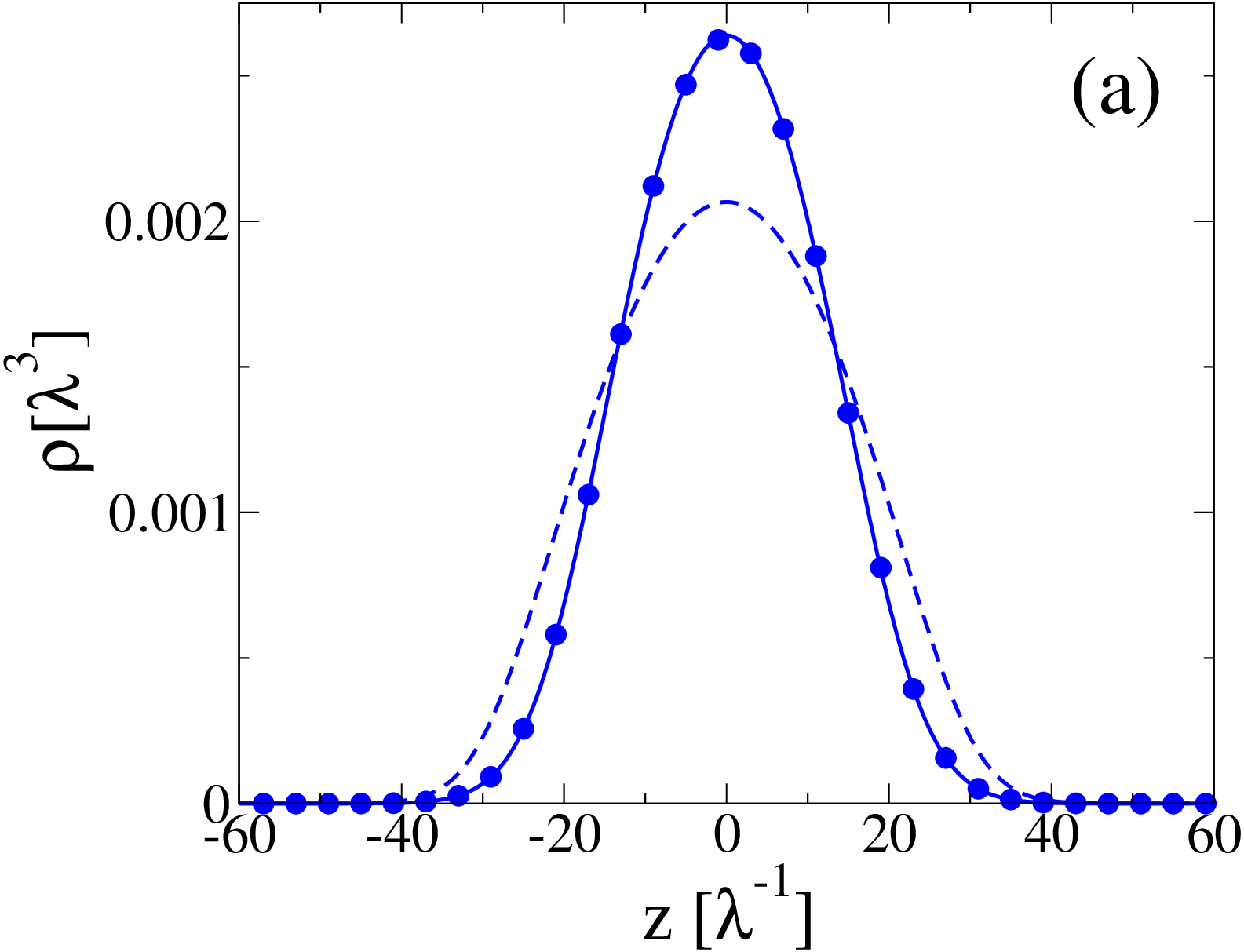}\vspace{0.4cm}
\includegraphics[width=8.cm]{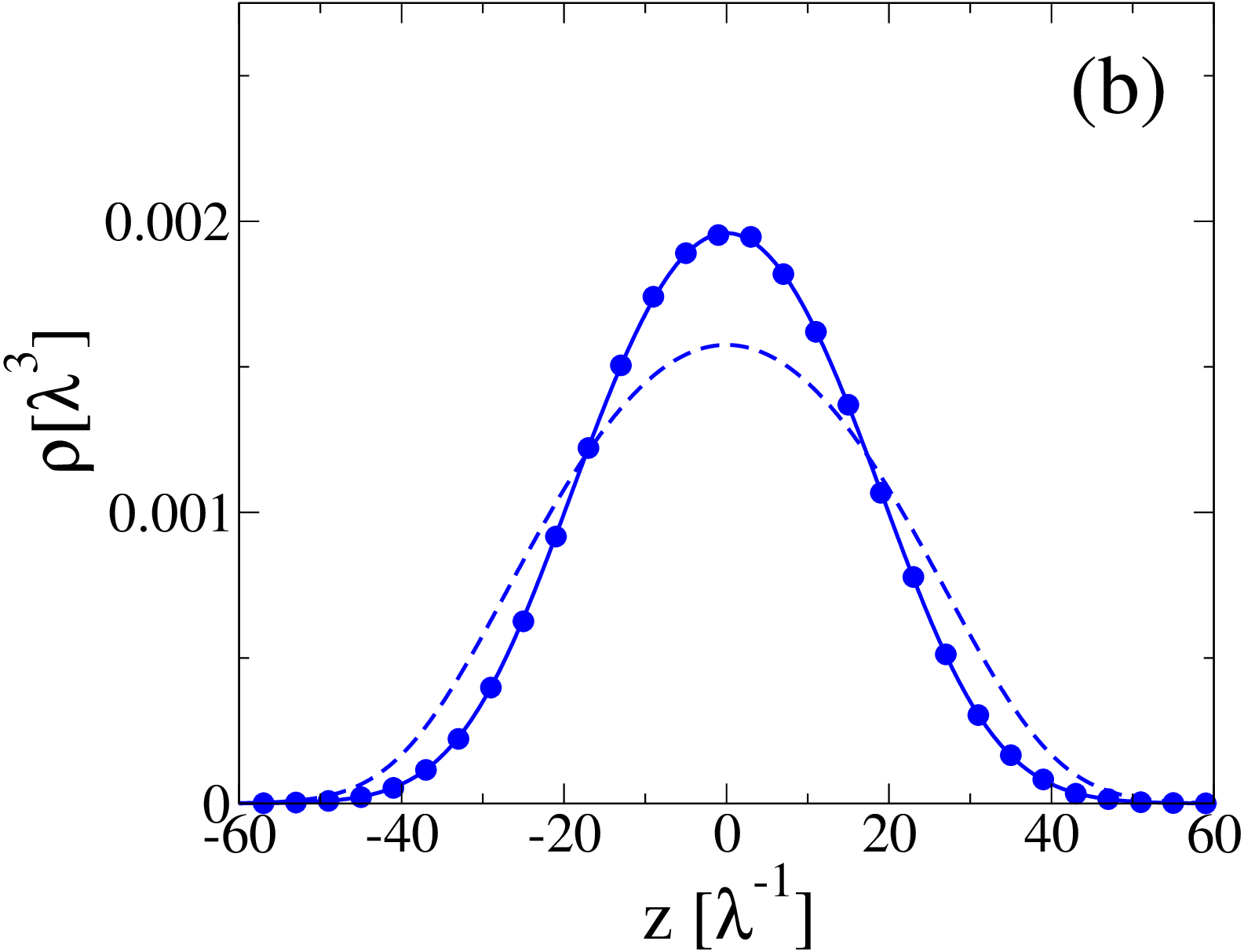}
\caption{Concentration profiles for $\epsilon=100$. The symbols represent MC simulation data, the solid curves represent the DFT, and the dashed curves the MF theory. The $\chi$ parameters are $0.01$ and $0.005$, for (a) and (b) figures, respectively.}
\label{fig3}
\end{figure}
\begin{figure}[t]
\includegraphics[width=8.cm]{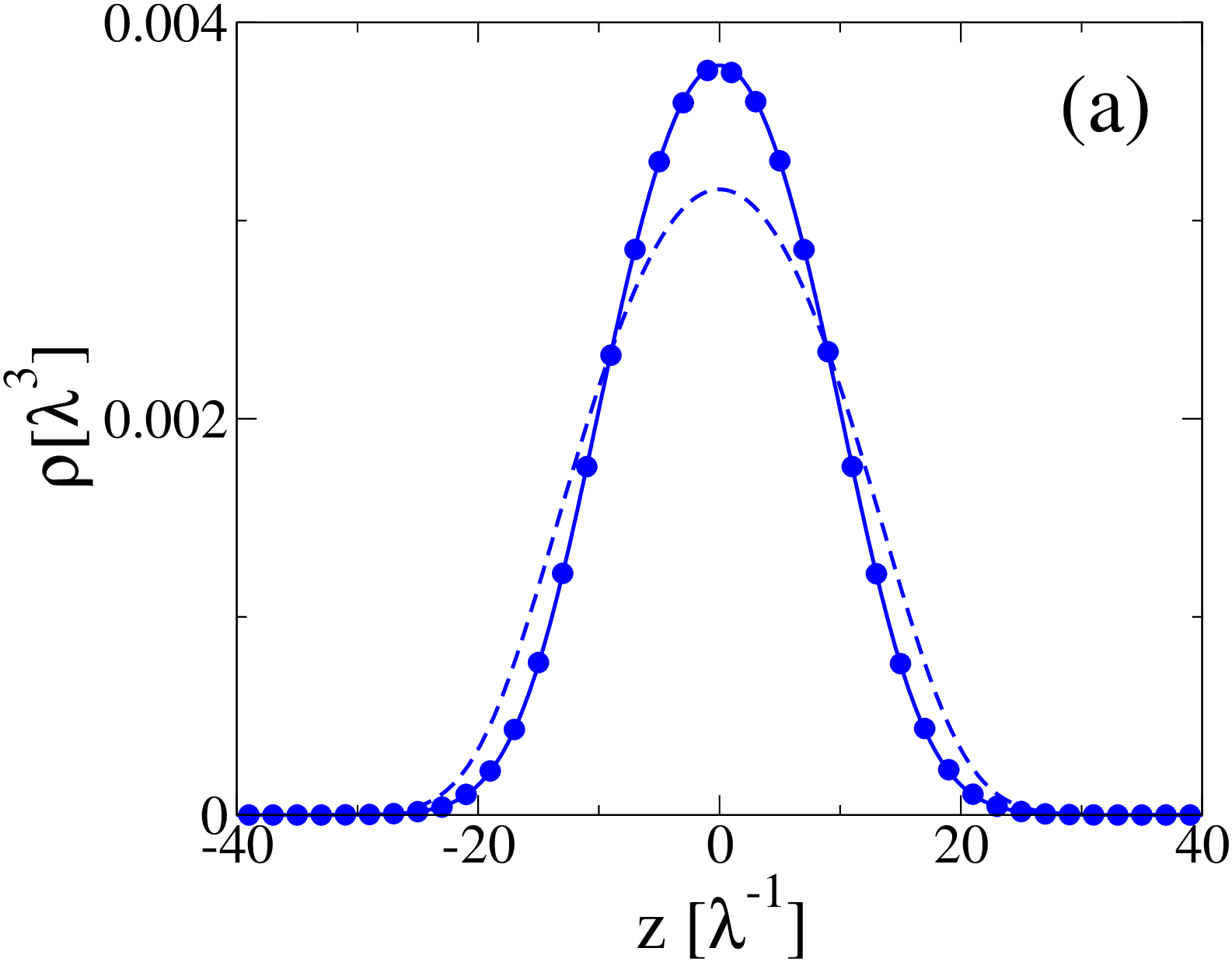}\vspace{0.4cm}
\includegraphics[width=8.cm]{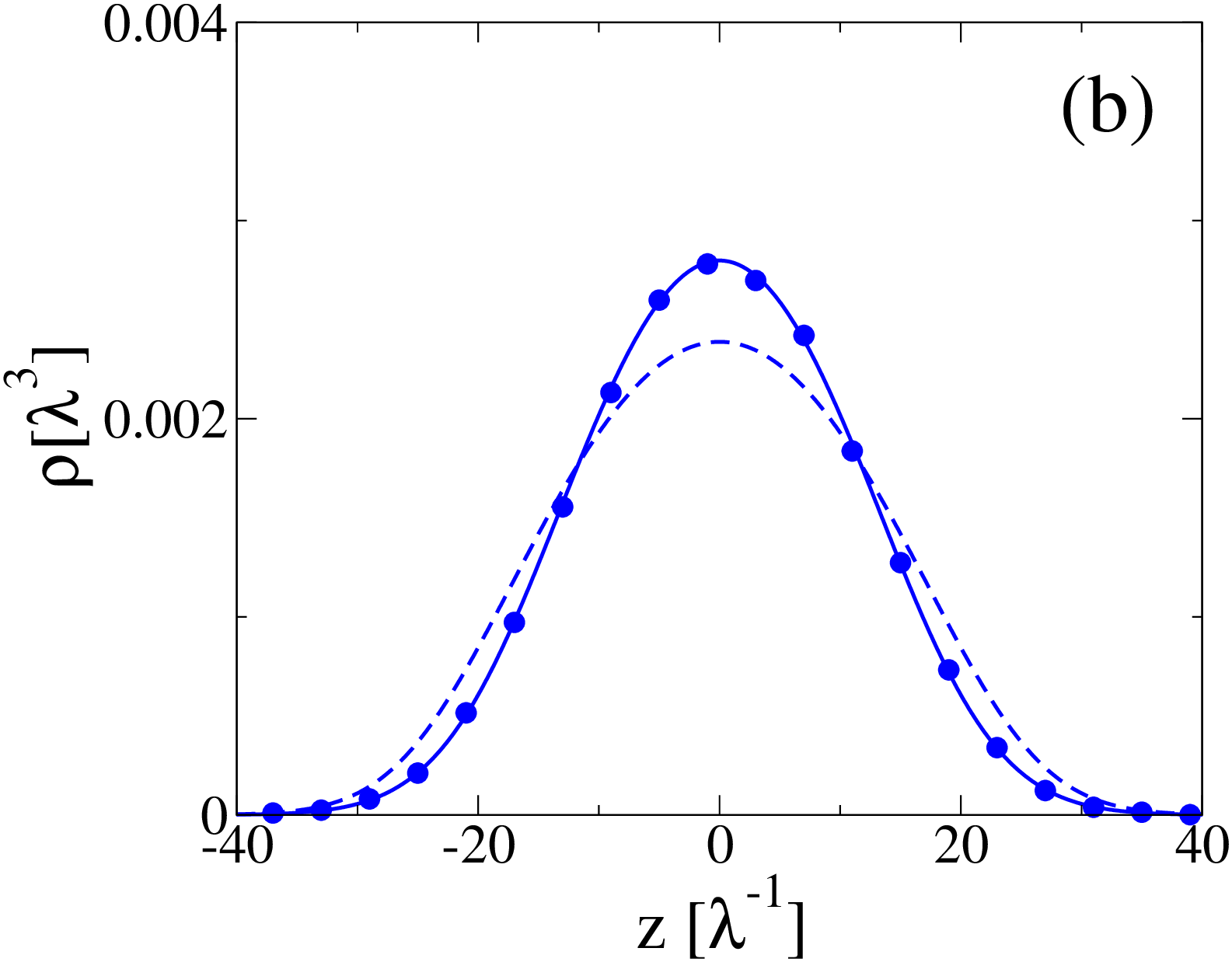}
\caption{Concentration profiles for $\epsilon=50$. The symbols represent MC simulation data, the solid curves represent the DFT, and the dashed curves the MF theory. The $\chi$ parameters are $0.02$ and $0.01$, for (a) and (b) figures, respectively.}
\label{fig4}
\end{figure}

To calculate the density profile of an inhomogeneous Yukawa gas confined by an external potential, the system of equations \eqref{dens2}, \eqref{phi2}, and the HNC equation must be solved simultaneously. In practice, to speed up the calculations, we first calculate the chemical potential of a homogeneous Yukawa system.  The solution of the HNC equation shows that to a very high degree of accuracy the correlational chemical potential has a simple parabolic form $\beta\mu^{cor}=a \rho + b \rho^2$.  The HNC equation  allows us to calculate the parameters $a$ and $b$ for various values of $\epsilon$.  To speed up the numerical integration,  we can then use the approximate form of the
LDA approximation, $\beta\mu^{cor}[\rho(z)]=a \rho(z) + b \rho^2(z)$, in Eq.~\eqref{dens2} .  In Figs.~\ref{fig3} and \ref{fig4}, we compare the theoretically calculated density profiles obtained using HNC-LDA with the results of  MC simulations.  We see that, while the MF theory fails to account for the simulation results, the DFT based on the HNC equation and the LDA is able
to provide an extremely accurate description of the particle distribution, without any adjustable parameters.  Perhaps surprisingly, the theory remain very accurate even in the very 
strong coupling limit of $\epsilon=100$. 

\section{Conclusions}

We have studied a gas of Yukawa particles confined by an external potential. In the weak coupling limit, we have constructed a MF theory which allows us to accurately calculate the equilibrium particle density distribution inside a confining potential. In the strong coupling limit, the correlations between the particles become important and the MF theory fails. We show, however, that a DFT theory based on the HNC equation and a LDA approximation accounts perfectly for the
observed particle distributions even in the limit of very strong interactions between the particles.  

\section{Acknowledgments}
This work was partially supported by the CNPq, INCT-FCx, and by the US-AFOSR under the grant 
FA9550-12-1-0438.


\begin{thebibliography}{49}%
\makeatletter
\providecommand \@ifxundefined [1]{%
 \@ifx{#1\undefined}
}%
\providecommand \@ifnum [1]{%
 \ifnum #1\expandafter \@firstoftwo
 \else \expandafter \@secondoftwo
 \fi
}%
\providecommand \@ifx [1]{%
 \ifx #1\expandafter \@firstoftwo
 \else \expandafter \@secondoftwo
 \fi
}%
\providecommand \natexlab [1]{#1}%
\providecommand \enquote  [1]{``#1''}%
\providecommand \bibnamefont  [1]{#1}%
\providecommand \bibfnamefont [1]{#1}%
\providecommand \citenamefont [1]{#1}%
\providecommand \href@noop [0]{\@secondoftwo}%
\providecommand \href [0]{\begingroup \@sanitize@url \@href}%
\providecommand \@href[1]{\@@startlink{#1}\@@href}%
\providecommand \@@href[1]{\endgroup#1\@@endlink}%
\providecommand \@sanitize@url [0]{\catcode `\\12\catcode `\$12\catcode
  `\&12\catcode `\#12\catcode `\^12\catcode `\_12\catcode `\%12\relax}%
\providecommand \@@startlink[1]{}%
\providecommand \@@endlink[0]{}%
\providecommand \url  [0]{\begingroup\@sanitize@url \@url }%
\providecommand \@url [1]{\endgroup\@href {#1}{\urlprefix }}%
\providecommand \urlprefix  [0]{URL }%
\providecommand \Eprint [0]{\href }%
\providecommand \doibase [0]{http://dx.doi.org/}%
\providecommand \selectlanguage [0]{\@gobble}%
\providecommand \bibinfo  [0]{\@secondoftwo}%
\providecommand \bibfield  [0]{\@secondoftwo}%
\providecommand \translation [1]{[#1]}%
\providecommand \BibitemOpen [0]{}%
\providecommand \bibitemStop [0]{}%
\providecommand \bibitemNoStop [0]{.\EOS\space}%
\providecommand \EOS [0]{\spacefactor3000\relax}%
\providecommand \BibitemShut  [1]{\csname bibitem#1\endcsname}%
\let\auto@bib@innerbib\@empty
\bibitem [{\citenamefont {Zammit}, \citenamefont {Fursa},\ and\ \citenamefont
  {Bray}(2010)}]{ZaFu10}%
  \BibitemOpen
  \bibfield  {author} {\bibinfo {author} {\bibfnamefont {M.~C.}\ \bibnamefont
  {Zammit}}, \bibinfo {author} {\bibfnamefont {D.~V.}\ \bibnamefont {Fursa}}, \
  and\ \bibinfo {author} {\bibfnamefont {I.}~\bibnamefont {Bray}},\ }\href@noop
  {} {\bibfield  {journal} {\bibinfo  {journal} {Phys. Rev. A}\ }\textbf
  {\bibinfo {volume} {82}},\ \bibinfo {pages} {052705} (\bibinfo {year}
  {2010})}\BibitemShut {NoStop}%
\bibitem [{\citenamefont {Basu}(2010)}]{Ba10}%
  \BibitemOpen
  \bibfield  {author} {\bibinfo {author} {\bibfnamefont {A.}~\bibnamefont
  {Basu}},\ }\href@noop {} {\bibfield  {journal} {\bibinfo  {journal} {J. Phys.
  B: At. Mol. Opt. Phys.}\ }\textbf {\bibinfo {volume} {43}},\ \bibinfo {pages}
  {115202} (\bibinfo {year} {2010})}\BibitemShut {NoStop}%
\bibitem [{\citenamefont {Totsuji}\ \emph {et~al.}(1998)\citenamefont
  {Totsuji}, \citenamefont {Kishimoto}, \citenamefont {Totsuji},\ and\
  \citenamefont {Sasabe}}]{ToKi98}%
  \BibitemOpen
  \bibfield  {author} {\bibinfo {author} {\bibfnamefont {H.}~\bibnamefont
  {Totsuji}}, \bibinfo {author} {\bibfnamefont {T.}~\bibnamefont {Kishimoto}},
  \bibinfo {author} {\bibfnamefont {C.}~\bibnamefont {Totsuji}}, \ and\
  \bibinfo {author} {\bibfnamefont {T.}~\bibnamefont {Sasabe}},\ }\href@noop {}
  {\bibfield  {journal} {\bibinfo  {journal} {Phys. Rev. E}\ }\textbf {\bibinfo
  {volume} {58}},\ \bibinfo {pages} {7831} (\bibinfo {year}
  {1998})}\BibitemShut {NoStop}%
\bibitem [{\citenamefont {Fortov}\ \emph {et~al.}(2005)\citenamefont {Fortov},
  \citenamefont {Ivlev}, \citenamefont {Khrapak}, \citenamefont {Khrapak},\
  and\ \citenamefont {Morfill}}]{FoIv05}%
  \BibitemOpen
  \bibfield  {author} {\bibinfo {author} {\bibfnamefont {V.~E.}\ \bibnamefont
  {Fortov}}, \bibinfo {author} {\bibfnamefont {A.~V.}\ \bibnamefont {Ivlev}},
  \bibinfo {author} {\bibfnamefont {S.~A.}\ \bibnamefont {Khrapak}}, \bibinfo
  {author} {\bibfnamefont {A.~G.}\ \bibnamefont {Khrapak}}, \ and\ \bibinfo
  {author} {\bibfnamefont {G.~E.}\ \bibnamefont {Morfill}},\ }\href@noop {}
  {\bibfield  {journal} {\bibinfo  {journal} {Phys. Rep.}\ }\textbf {\bibinfo
  {volume} {421}},\ \bibinfo {pages} {1} (\bibinfo {year} {2005})}\BibitemShut
  {NoStop}%
\bibitem [{\citenamefont {Dzhumagulova}, \citenamefont {Ramazanov},\ and\
  \citenamefont {Masheeva}(2013)}]{DzRa13}%
  \BibitemOpen
  \bibfield  {author} {\bibinfo {author} {\bibfnamefont {K.~N.}\ \bibnamefont
  {Dzhumagulova}}, \bibinfo {author} {\bibfnamefont {T.~S.}\ \bibnamefont
  {Ramazanov}}, \ and\ \bibinfo {author} {\bibfnamefont {R.~U.}\ \bibnamefont
  {Masheeva}},\ }\href@noop {} {\bibfield  {journal} {\bibinfo  {journal}
  {Phys. Plasmas}\ }\textbf {\bibinfo {volume} {20}},\ \bibinfo {pages}
  {113702} (\bibinfo {year} {2013})}\BibitemShut {NoStop}%
\bibitem [{\citenamefont {Meijer}\ and\ \citenamefont
  {Frenkel}(1991{\natexlab{a}})}]{MeFr91b}%
  \BibitemOpen
  \bibfield  {author} {\bibinfo {author} {\bibfnamefont {E.~J.}\ \bibnamefont
  {Meijer}}\ and\ \bibinfo {author} {\bibfnamefont {D.}~\bibnamefont
  {Frenkel}},\ }\href@noop {} {\bibfield  {journal} {\bibinfo  {journal} {Phys.
  Rev. Lett.}\ }\textbf {\bibinfo {volume} {67}},\ \bibinfo {pages} {1110}
  (\bibinfo {year} {1991}{\natexlab{a}})}\BibitemShut {NoStop}%
\bibitem [{\citenamefont {Heinen}\ \emph {et~al.}(2011)\citenamefont {Heinen},
  \citenamefont {Holmqvist}, \citenamefont {Banchio},\ and\ \citenamefont
  {N\"agele}}]{HeHo11}%
  \BibitemOpen
  \bibfield  {author} {\bibinfo {author} {\bibfnamefont {M.}~\bibnamefont
  {Heinen}}, \bibinfo {author} {\bibfnamefont {P.}~\bibnamefont {Holmqvist}},
  \bibinfo {author} {\bibfnamefont {A.~J.}\ \bibnamefont {Banchio}}, \ and\
  \bibinfo {author} {\bibfnamefont {G.}~\bibnamefont {N\"agele}},\ }\href@noop
  {} {\bibfield  {journal} {\bibinfo  {journal} {J. Chem. Phys.}\ }\textbf
  {\bibinfo {volume} {134}},\ \bibinfo {pages} {044532} (\bibinfo {year}
  {2011})}\BibitemShut {NoStop}%
\bibitem [{\citenamefont {{van der Linden}}, \citenamefont {{van Blaaderen}},\
  and\ \citenamefont {Dijkstra}(2013)}]{vava13}%
  \BibitemOpen
  \bibfield  {author} {\bibinfo {author} {\bibfnamefont {M.~N.}\ \bibnamefont
  {{van der Linden}}}, \bibinfo {author} {\bibfnamefont {A.}~\bibnamefont {{van
  Blaaderen}}}, \ and\ \bibinfo {author} {\bibfnamefont {M.}~\bibnamefont
  {Dijkstra}},\ }\href@noop {} {\bibfield  {journal} {\bibinfo  {journal} {J.
  Chem. Phys.}\ }\textbf {\bibinfo {volume} {138}},\ \bibinfo {pages} {114903}
  (\bibinfo {year} {2013})}\BibitemShut {NoStop}%
\bibitem [{\citenamefont {Jiao}\ and\ \citenamefont {Ho}(2013)}]{JiHo13}%
  \BibitemOpen
  \bibfield  {author} {\bibinfo {author} {\bibfnamefont {L.~G.}\ \bibnamefont
  {Jiao}}\ and\ \bibinfo {author} {\bibfnamefont {Y.~K.}\ \bibnamefont {Ho}},\
  }\href@noop {} {\bibfield  {journal} {\bibinfo  {journal} {Int. J. Quantum
  Chem.}\ }\textbf {\bibinfo {volume} {113}},\ \bibinfo {pages} {2569}
  (\bibinfo {year} {2013})}\BibitemShut {NoStop}%
\bibitem [{\citenamefont {Certik}\ and\ \citenamefont
  {Winkler}(2013)}]{CeWi13}%
  \BibitemOpen
  \bibfield  {author} {\bibinfo {author} {\bibfnamefont {O.}~\bibnamefont
  {Certik}}\ and\ \bibinfo {author} {\bibfnamefont {P.}~\bibnamefont
  {Winkler}},\ }\href@noop {} {\bibfield  {journal} {\bibinfo  {journal} {Int.
  J. Quantum Chem.}\ }\textbf {\bibinfo {volume} {113}},\ \bibinfo {pages}
  {2012} (\bibinfo {year} {2013})}\BibitemShut {NoStop}%
\bibitem [{\citenamefont {Likos}(2001)}]{Li01}%
  \BibitemOpen
  \bibfield  {author} {\bibinfo {author} {\bibfnamefont {C.~N.}\ \bibnamefont
  {Likos}},\ }\href@noop {} {\bibfield  {journal} {\bibinfo  {journal} {Phys.
  Rep.}\ }\textbf {\bibinfo {volume} {348}},\ \bibinfo {pages} {247} (\bibinfo
  {year} {2001})}\BibitemShut {NoStop}%
\bibitem [{\citenamefont {Levin}(2002)}]{Le02}%
  \BibitemOpen
  \bibfield  {author} {\bibinfo {author} {\bibfnamefont {Y.}~\bibnamefont
  {Levin}},\ }\href@noop {} {\bibfield  {journal} {\bibinfo  {journal} {Rep.
  Prog. Phys.}\ }\textbf {\bibinfo {volume} {65}},\ \bibinfo {pages} {1577}
  (\bibinfo {year} {2002})}\BibitemShut {NoStop}%
\bibitem [{\citenamefont {Hamaguchi}, \citenamefont {Farouki},\ and\
  \citenamefont {Dubin}(1991)}]{HaFa91}%
  \BibitemOpen
  \bibfield  {author} {\bibinfo {author} {\bibfnamefont {S.}~\bibnamefont
  {Hamaguchi}}, \bibinfo {author} {\bibfnamefont {R.~T.}\ \bibnamefont
  {Farouki}}, \ and\ \bibinfo {author} {\bibfnamefont {D.~H.~E.}\ \bibnamefont
  {Dubin}},\ }\href@noop {} {\bibfield  {journal} {\bibinfo  {journal} {Phys.
  Rev. E}\ }\textbf {\bibinfo {volume} {56}},\ \bibinfo {pages} {4671}
  (\bibinfo {year} {1991})}\BibitemShut {NoStop}%
\bibitem [{\citenamefont {Hagen}\ and\ \citenamefont {Frenkel}(1994)}]{HaFr94}%
  \BibitemOpen
  \bibfield  {author} {\bibinfo {author} {\bibfnamefont {M.~H.~J.}\
  \bibnamefont {Hagen}}\ and\ \bibinfo {author} {\bibfnamefont
  {D.}~\bibnamefont {Frenkel}},\ }\href@noop {} {\bibfield  {journal} {\bibinfo
   {journal} {J. Chem. Phys.}\ }\textbf {\bibinfo {volume} {101}},\ \bibinfo
  {pages} {4093} (\bibinfo {year} {1994})}\BibitemShut {NoStop}%
\bibitem [{\citenamefont {Scholl-Paschinger}\ \emph {et~al.}(2013)\citenamefont
  {Scholl-Paschinger}, \citenamefont {Valadez-Perez}, \citenamefont
  {Benavides},\ and\ \citenamefont {Castaneda-Priego}}]{ScVa13}%
  \BibitemOpen
  \bibfield  {author} {\bibinfo {author} {\bibfnamefont {E.}~\bibnamefont
  {Scholl-Paschinger}}, \bibinfo {author} {\bibfnamefont {N.~E.}\ \bibnamefont
  {Valadez-Perez}}, \bibinfo {author} {\bibfnamefont {A.~L.}\ \bibnamefont
  {Benavides}}, \ and\ \bibinfo {author} {\bibfnamefont {R.}~\bibnamefont
  {Castaneda-Priego}},\ }\href@noop {} {\bibfield  {journal} {\bibinfo
  {journal} {J. Chem. Phys.}\ }\textbf {\bibinfo {volume} {139}},\ \bibinfo
  {pages} {184902} (\bibinfo {year} {2013})}\BibitemShut {NoStop}%
\bibitem [{\citenamefont {Robbins}, \citenamefont {Kremer},\ and\ \citenamefont
  {Grest}(1988)}]{RoKr88}%
  \BibitemOpen
  \bibfield  {author} {\bibinfo {author} {\bibfnamefont {M.~O.}\ \bibnamefont
  {Robbins}}, \bibinfo {author} {\bibfnamefont {K.}~\bibnamefont {Kremer}}, \
  and\ \bibinfo {author} {\bibfnamefont {G.~S.}\ \bibnamefont {Grest}},\
  }\href@noop {} {\bibfield  {journal} {\bibinfo  {journal} {J. Chem. Phys.}\
  }\textbf {\bibinfo {volume} {88}},\ \bibinfo {pages} {3286} (\bibinfo {year}
  {1988})}\BibitemShut {NoStop}%
\bibitem [{\citenamefont {Meijer}\ and\ \citenamefont
  {Frenkel}(1991{\natexlab{b}})}]{MeFr91a}%
  \BibitemOpen
  \bibfield  {author} {\bibinfo {author} {\bibfnamefont {E.~J.}\ \bibnamefont
  {Meijer}}\ and\ \bibinfo {author} {\bibfnamefont {D.}~\bibnamefont
  {Frenkel}},\ }\href@noop {} {\bibfield  {journal} {\bibinfo  {journal} {J.
  Chem. Phys.}\ }\textbf {\bibinfo {volume} {94}},\ \bibinfo {pages} {2269}
  (\bibinfo {year} {1991}{\natexlab{b}})}\BibitemShut {NoStop}%
\bibitem [{\citenamefont {L\"owen}, \citenamefont {Palberg},\ and\
  \citenamefont {Simon}(1993)}]{LoPa93}%
  \BibitemOpen
  \bibfield  {author} {\bibinfo {author} {\bibfnamefont {H.}~\bibnamefont
  {L\"owen}}, \bibinfo {author} {\bibfnamefont {T.}~\bibnamefont {Palberg}}, \
  and\ \bibinfo {author} {\bibfnamefont {R.}~\bibnamefont {Simon}},\
  }\href@noop {} {\bibfield  {journal} {\bibinfo  {journal} {Phys. Rev. Lett.}\
  }\textbf {\bibinfo {volume} {70}},\ \bibinfo {pages} {1557} (\bibinfo {year}
  {1993})}\BibitemShut {NoStop}%
\bibitem [{\citenamefont {Palberg}\ \emph {et~al.}(1995)\citenamefont
  {Palberg}, \citenamefont {M\"onch}, \citenamefont {Bitzer}, \citenamefont
  {Bellini},\ and\ \citenamefont {Piazza}}]{PaMo95}%
  \BibitemOpen
  \bibfield  {author} {\bibinfo {author} {\bibfnamefont {T.}~\bibnamefont
  {Palberg}}, \bibinfo {author} {\bibfnamefont {W.}~\bibnamefont {M\"onch}},
  \bibinfo {author} {\bibfnamefont {F.}~\bibnamefont {Bitzer}}, \bibinfo
  {author} {\bibfnamefont {T.}~\bibnamefont {Bellini}}, \ and\ \bibinfo
  {author} {\bibfnamefont {R.}~\bibnamefont {Piazza}},\ }\href@noop {}
  {\bibfield  {journal} {\bibinfo  {journal} {Phys. Rev. Lett.}\ }\textbf
  {\bibinfo {volume} {74}},\ \bibinfo {pages} {4555} (\bibinfo {year}
  {1995})}\BibitemShut {NoStop}%
\bibitem [{\citenamefont {Stevens}\ and\ \citenamefont
  {Robbins}(1998)}]{StRo98}%
  \BibitemOpen
  \bibfield  {author} {\bibinfo {author} {\bibfnamefont {M.~J.}\ \bibnamefont
  {Stevens}}\ and\ \bibinfo {author} {\bibfnamefont {M.~O.}\ \bibnamefont
  {Robbins}},\ }\href@noop {} {\bibfield  {journal} {\bibinfo  {journal} {J.
  Chem. Phys.}\ }\textbf {\bibinfo {volume} {98}},\ \bibinfo {pages} {2319}
  (\bibinfo {year} {1998})}\BibitemShut {NoStop}%
\bibitem [{\citenamefont {Hoy}\ and\ \citenamefont {Robbins}(2004)}]{HoRo04}%
  \BibitemOpen
  \bibfield  {author} {\bibinfo {author} {\bibfnamefont {R.~S.}\ \bibnamefont
  {Hoy}}\ and\ \bibinfo {author} {\bibfnamefont {M.~O.}\ \bibnamefont
  {Robbins}},\ }\href@noop {} {\bibfield  {journal} {\bibinfo  {journal} {Phys.
  Rev. E}\ }\textbf {\bibinfo {volume} {69}},\ \bibinfo {pages} {056103}
  (\bibinfo {year} {2004})}\BibitemShut {NoStop}%
\bibitem [{\citenamefont {Gapinski}, \citenamefont {N\"agele},\ and\
  \citenamefont {Patkowski}(2012)}]{GaNa12}%
  \BibitemOpen
  \bibfield  {author} {\bibinfo {author} {\bibfnamefont {J.}~\bibnamefont
  {Gapinski}}, \bibinfo {author} {\bibfnamefont {G.}~\bibnamefont {N\"agele}},
  \ and\ \bibinfo {author} {\bibfnamefont {A.}~\bibnamefont {Patkowski}},\
  }\href@noop {} {\bibfield  {journal} {\bibinfo  {journal} {J. Chem. Phys.}\
  }\textbf {\bibinfo {volume} {136}},\ \bibinfo {pages} {024507} (\bibinfo
  {year} {2012})}\BibitemShut {NoStop}%
\bibitem [{\citenamefont {Crocker}\ and\ \citenamefont {Grier}(1996)}]{CrGr96}%
  \BibitemOpen
  \bibfield  {author} {\bibinfo {author} {\bibfnamefont {J.~C.}\ \bibnamefont
  {Crocker}}\ and\ \bibinfo {author} {\bibfnamefont {D.~G.}\ \bibnamefont
  {Grier}},\ }\href@noop {} {\bibfield  {journal} {\bibinfo  {journal} {Phys.
  Rev. Lett.}\ }\textbf {\bibinfo {volume} {77}},\ \bibinfo {pages} {1897}
  (\bibinfo {year} {1996})}\BibitemShut {NoStop}%
\bibitem [{\citenamefont {Crocker}(1997)}]{Cr97}%
  \BibitemOpen
  \bibfield  {author} {\bibinfo {author} {\bibfnamefont {J.~C.}\ \bibnamefont
  {Crocker}},\ }\href@noop {} {\bibfield  {journal} {\bibinfo  {journal} {J.
  Chem. Phys.}\ }\textbf {\bibinfo {volume} {106}},\ \bibinfo {pages} {2837}
  (\bibinfo {year} {1997})}\BibitemShut {NoStop}%
\bibitem [{\citenamefont {Grier}(1997)}]{Gr97}%
  \BibitemOpen
  \bibfield  {author} {\bibinfo {author} {\bibfnamefont {D.~G.}\ \bibnamefont
  {Grier}},\ }\href@noop {} {\bibfield  {journal} {\bibinfo  {journal} {Curr.
  Opin. Colloid Interface Sci.}\ }\textbf {\bibinfo {volume} {2}},\ \bibinfo
  {pages} {264} (\bibinfo {year} {1997})}\BibitemShut {NoStop}%
\bibitem [{\citenamefont {Dufresne}\ and\ \citenamefont
  {Grier}(1998)}]{DuGr98}%
  \BibitemOpen
  \bibfield  {author} {\bibinfo {author} {\bibfnamefont {E.~R.}\ \bibnamefont
  {Dufresne}}\ and\ \bibinfo {author} {\bibfnamefont {D.~G.}\ \bibnamefont
  {Grier}},\ }\href@noop {} {\bibfield  {journal} {\bibinfo  {journal} {Rev.
  Sci. Instrum.}\ }\textbf {\bibinfo {volume} {69}},\ \bibinfo {pages} {1974}
  (\bibinfo {year} {1998})}\BibitemShut {NoStop}%
\bibitem [{\citenamefont {Lin}, \citenamefont {Yu},\ and\ \citenamefont
  {Rice}(2000)}]{LiYu00}%
  \BibitemOpen
  \bibfield  {author} {\bibinfo {author} {\bibfnamefont {B.~H.}\ \bibnamefont
  {Lin}}, \bibinfo {author} {\bibfnamefont {J.}~\bibnamefont {Yu}}, \ and\
  \bibinfo {author} {\bibfnamefont {S.~A.}\ \bibnamefont {Rice}},\ }\href@noop
  {} {\bibfield  {journal} {\bibinfo  {journal} {Phys. Rev. E}\ }\textbf
  {\bibinfo {volume} {62}},\ \bibinfo {pages} {3909} (\bibinfo {year}
  {2000})}\BibitemShut {NoStop}%
\bibitem [{\citenamefont {L\"owen}(2001)}]{Lo01}%
  \BibitemOpen
  \bibfield  {author} {\bibinfo {author} {\bibfnamefont {H.}~\bibnamefont
  {L\"owen}},\ }\href@noop {} {\bibfield  {journal} {\bibinfo  {journal} {J.
  Phys.: Condens. Matter}\ }\textbf {\bibinfo {volume} {13}},\ \bibinfo {pages}
  {R415} (\bibinfo {year} {2001})}\BibitemShut {NoStop}%
\bibitem [{\citenamefont {Resnick}(2003)}]{Re03}%
  \BibitemOpen
  \bibfield  {author} {\bibinfo {author} {\bibfnamefont {A.}~\bibnamefont
  {Resnick}},\ }\href@noop {} {\bibfield  {journal} {\bibinfo  {journal} {J.
  Colloid Interface Sci.}\ }\textbf {\bibinfo {volume} {262}},\ \bibinfo
  {pages} {55} (\bibinfo {year} {2003})}\BibitemShut {NoStop}%
\bibitem [{\citenamefont {Elmahdy}, \citenamefont {Gutsche},\ and\
  \citenamefont {Kremer}(2010)}]{ElGu10}%
  \BibitemOpen
  \bibfield  {author} {\bibinfo {author} {\bibfnamefont {M.~M.}\ \bibnamefont
  {Elmahdy}}, \bibinfo {author} {\bibfnamefont {C.}~\bibnamefont {Gutsche}}, \
  and\ \bibinfo {author} {\bibfnamefont {F.}~\bibnamefont {Kremer}},\
  }\href@noop {} {\bibfield  {journal} {\bibinfo  {journal} {J. Phys. Chem. C}\
  }\textbf {\bibinfo {volume} {114}},\ \bibinfo {pages} {19452} (\bibinfo
  {year} {2010})}\BibitemShut {NoStop}%
\bibitem [{\citenamefont {Levin}\ and\ \citenamefont {Pakter}(2011)}]{LePa11}%
  \BibitemOpen
  \bibfield  {author} {\bibinfo {author} {\bibfnamefont {Y.}~\bibnamefont
  {Levin}}\ and\ \bibinfo {author} {\bibfnamefont {R.}~\bibnamefont {Pakter}},\
  }\href@noop {} {\bibfield  {journal} {\bibinfo  {journal} {Phys. Rev. Lett.}\
  }\textbf {\bibinfo {volume} {107}},\ \bibinfo {pages} {088901} (\bibinfo
  {year} {2011})}\BibitemShut {NoStop}%
\bibitem [{\citenamefont {Girotto}, \citenamefont {{dos Santos}},\ and\
  \citenamefont {Levin}(2013)}]{GiDo13}%
  \BibitemOpen
  \bibfield  {author} {\bibinfo {author} {\bibfnamefont {M.}~\bibnamefont
  {Girotto}}, \bibinfo {author} {\bibfnamefont {A.~P.}\ \bibnamefont {{dos
  Santos}}}, \ and\ \bibinfo {author} {\bibfnamefont {Y.}~\bibnamefont
  {Levin}},\ }\href@noop {} {\bibfield  {journal} {\bibinfo  {journal} {Phys.
  Rev. E}\ }\textbf {\bibinfo {volume} {88}},\ \bibinfo {pages} {032118}
  (\bibinfo {year} {2013})}\BibitemShut {NoStop}%
\bibitem [{\citenamefont {van Roij}\ and\ \citenamefont
  {Hansen}(1997)}]{RoHa97}%
  \BibitemOpen
  \bibfield  {author} {\bibinfo {author} {\bibfnamefont {R.}~\bibnamefont {van
  Roij}}\ and\ \bibinfo {author} {\bibfnamefont {J.~P.}\ \bibnamefont
  {Hansen}},\ }\href@noop {} {\bibfield  {journal} {\bibinfo  {journal} {Phys.
  Rev. Lett.}\ }\textbf {\bibinfo {volume} {79}},\ \bibinfo {pages} {3082}
  (\bibinfo {year} {1997})}\BibitemShut {NoStop}%
\bibitem [{\citenamefont {Alexander}\ \emph {et~al.}(1984)\citenamefont
  {Alexander}, \citenamefont {Chaikin}, \citenamefont {Grant}, \citenamefont
  {Morales}, \citenamefont {Pincus},\ and\ \citenamefont {Hone}}]{AlCh84}%
  \BibitemOpen
  \bibfield  {author} {\bibinfo {author} {\bibfnamefont {S.}~\bibnamefont
  {Alexander}}, \bibinfo {author} {\bibfnamefont {P.~M.}\ \bibnamefont
  {Chaikin}}, \bibinfo {author} {\bibfnamefont {P.}~\bibnamefont {Grant}},
  \bibinfo {author} {\bibfnamefont {G.~J.}\ \bibnamefont {Morales}}, \bibinfo
  {author} {\bibfnamefont {P.}~\bibnamefont {Pincus}}, \ and\ \bibinfo {author}
  {\bibfnamefont {D.}~\bibnamefont {Hone}},\ }\href@noop {} {\bibfield
  {journal} {\bibinfo  {journal} {J. Chem. Phys.}\ }\textbf {\bibinfo {volume}
  {80}},\ \bibinfo {pages} {5776} (\bibinfo {year} {1984})}\BibitemShut
  {NoStop}%
\bibitem [{\citenamefont {Levin}, \citenamefont {Barbosa},\ and\ \citenamefont
  {Diehl}(1998)}]{Le98}%
  \BibitemOpen
  \bibfield  {author} {\bibinfo {author} {\bibfnamefont {Y.}~\bibnamefont
  {Levin}}, \bibinfo {author} {\bibfnamefont {M.~C.}\ \bibnamefont {Barbosa}},
  \ and\ \bibinfo {author} {\bibfnamefont {A.}~\bibnamefont {Diehl}},\
  }\href@noop {} {\bibfield  {journal} {\bibinfo  {journal} {Europhys. Lett.}\
  }\textbf {\bibinfo {volume} {41}},\ \bibinfo {pages} {123} (\bibinfo {year}
  {1998})}\BibitemShut {NoStop}%
\bibitem [{\citenamefont {Grosber}, \citenamefont {Nguyen},\ and\ \citenamefont
  {Shklovskii}(2002)}]{Sh02}%
  \BibitemOpen
  \bibfield  {author} {\bibinfo {author} {\bibfnamefont {A.~Y.}\ \bibnamefont
  {Grosber}}, \bibinfo {author} {\bibfnamefont {T.~T.}\ \bibnamefont {Nguyen}},
  \ and\ \bibinfo {author} {\bibfnamefont {B.~I.}\ \bibnamefont {Shklovskii}},\
  }\href@noop {} {\bibfield  {journal} {\bibinfo  {journal} {Rev. Mod. Phys.}\
  }\textbf {\bibinfo {volume} {74}},\ \bibinfo {pages} {329} (\bibinfo {year}
  {2002})}\BibitemShut {NoStop}%
\bibitem [{\citenamefont {Pianegonda}, \citenamefont {Barbosa},\ and\
  \citenamefont {Levin}(2005)}]{PiBa05}%
  \BibitemOpen
  \bibfield  {author} {\bibinfo {author} {\bibfnamefont {S.}~\bibnamefont
  {Pianegonda}}, \bibinfo {author} {\bibfnamefont {M.~C.}\ \bibnamefont
  {Barbosa}}, \ and\ \bibinfo {author} {\bibfnamefont {Y.}~\bibnamefont
  {Levin}},\ }\href@noop {} {\bibfield  {journal} {\bibinfo  {journal}
  {Europhys. Lett.}\ }\textbf {\bibinfo {volume} {71}},\ \bibinfo {pages} {831}
  (\bibinfo {year} {2005})}\BibitemShut {NoStop}%
\bibitem [{\citenamefont {dos Santos}, \citenamefont {Diehl},\ and\
  \citenamefont {Levin}(2010)}]{DoLe10}%
  \BibitemOpen
  \bibfield  {author} {\bibinfo {author} {\bibfnamefont {A.~P.}\ \bibnamefont
  {dos Santos}}, \bibinfo {author} {\bibfnamefont {A.}~\bibnamefont {Diehl}}, \
  and\ \bibinfo {author} {\bibfnamefont {Y.}~\bibnamefont {Levin}},\
  }\href@noop {} {\bibfield  {journal} {\bibinfo  {journal} {J. Chem. Phys.}\
  }\textbf {\bibinfo {volume} {132}},\ \bibinfo {pages} {104105} (\bibinfo
  {year} {2010})}\BibitemShut {NoStop}%
\bibitem [{\citenamefont {Rosenfeld}(1989)}]{Ro89}%
  \BibitemOpen
  \bibfield  {author} {\bibinfo {author} {\bibfnamefont {Y.}~\bibnamefont
  {Rosenfeld}},\ }\href@noop {} {\bibfield  {journal} {\bibinfo  {journal}
  {Phys. Rev. Lett.}\ }\textbf {\bibinfo {volume} {63}},\ \bibinfo {pages}
  {980} (\bibinfo {year} {1989})}\BibitemShut {NoStop}%
\bibitem [{\citenamefont {Rosenfeld}\ \emph {et~al.}(1996)\citenamefont
  {Rosenfeld}, \citenamefont {Schmidt}, \citenamefont {L\"owen},\ and\
  \citenamefont {Tarazona}}]{RoSc96}%
  \BibitemOpen
  \bibfield  {author} {\bibinfo {author} {\bibfnamefont {Y.}~\bibnamefont
  {Rosenfeld}}, \bibinfo {author} {\bibfnamefont {M.}~\bibnamefont {Schmidt}},
  \bibinfo {author} {\bibfnamefont {H.}~\bibnamefont {L\"owen}}, \ and\
  \bibinfo {author} {\bibfnamefont {P.}~\bibnamefont {Tarazona}},\ }\href@noop
  {} {\bibfield  {journal} {\bibinfo  {journal} {J. Phys.: Condens. Matter}\
  }\textbf {\bibinfo {volume} {8}},\ \bibinfo {pages} {L577} (\bibinfo {year}
  {1996})}\BibitemShut {NoStop}%
\bibitem [{\citenamefont {{R Evans}}(2009)}]{Ev09}%
  \BibitemOpen
  \bibfield  {author} {\bibinfo {author} {\bibnamefont {{R Evans}}},\
  }\href@noop {} {\emph {\bibinfo {title} {Lecture Notes at 3rd Warsaw School
  of Statistical Physics}}}\ (\bibinfo  {publisher} {Warsaw University Press},\
  \bibinfo {address} {Kazimierz Dolny},\ \bibinfo {year} {2009})\ pp.\ \bibinfo
  {pages} {43--85}\BibitemShut {NoStop}%
\bibitem [{\citenamefont {Roth}(2010)}]{Ro10}%
  \BibitemOpen
  \bibfield  {author} {\bibinfo {author} {\bibfnamefont {R.}~\bibnamefont
  {Roth}},\ }\href@noop {} {\bibfield  {journal} {\bibinfo  {journal} {J.
  Phys.: Condens. Matter}\ }\textbf {\bibinfo {volume} {22}},\ \bibinfo {pages}
  {063102} (\bibinfo {year} {2010})}\BibitemShut {NoStop}%
\bibitem [{\citenamefont {Frydel}\ and\ \citenamefont {Levin}(2013)}]{FrLe13}%
  \BibitemOpen
  \bibfield  {author} {\bibinfo {author} {\bibfnamefont {D.}~\bibnamefont
  {Frydel}}\ and\ \bibinfo {author} {\bibfnamefont {Y.}~\bibnamefont {Levin}},\
  }\href@noop {} {\bibfield  {journal} {\bibinfo  {journal} {J. Chem. Phys.}\
  }\textbf {\bibinfo {volume} {138}},\ \bibinfo {pages} {174901} (\bibinfo
  {year} {2013})}\BibitemShut {NoStop}%
\bibitem [{\citenamefont {Evans}(1979)}]{Ev79}%
  \BibitemOpen
  \bibfield  {author} {\bibinfo {author} {\bibfnamefont {R.}~\bibnamefont
  {Evans}},\ }\href@noop {} {\bibfield  {journal} {\bibinfo  {journal} {Adv.
  Phys.}\ }\textbf {\bibinfo {volume} {28}},\ \bibinfo {pages} {143} (\bibinfo
  {year} {1979})}\BibitemShut {NoStop}%
\bibitem [{\citenamefont {{J. P. Hansen and I. R. McDonald}}(2006)}]{HaMc06}%
  \BibitemOpen
  \bibfield  {author} {\bibinfo {author} {\bibnamefont {{J. P. Hansen and I. R.
  McDonald}}},\ }\href@noop {} {\emph {\bibinfo {title} {Theory of Simple
  Liquids}}},\ \bibinfo {edition} {3rd}\ ed.\ (\bibinfo  {publisher}
  {Academic},\ \bibinfo {address} {London},\ \bibinfo {year}
  {2006})\BibitemShut {NoStop}%
\bibitem [{\citenamefont {Attard}(1996)}]{At96}%
  \BibitemOpen
  \bibfield  {author} {\bibinfo {author} {\bibfnamefont {P.}~\bibnamefont
  {Attard}},\ }\href@noop {} {\bibfield  {journal} {\bibinfo  {journal} {Avd.
  Chem. Phys.}\ }\textbf {\bibinfo {volume} {92}},\ \bibinfo {pages} {1}
  (\bibinfo {year} {1996})}\BibitemShut {NoStop}%
\bibitem [{\citenamefont {Colla}, \citenamefont {{dos Santos}},\ and\
  \citenamefont {Levin}(2012)}]{CoDo12}%
  \BibitemOpen
  \bibfield  {author} {\bibinfo {author} {\bibfnamefont {T.~E.}\ \bibnamefont
  {Colla}}, \bibinfo {author} {\bibfnamefont {A.~P.}\ \bibnamefont {{dos
  Santos}}}, \ and\ \bibinfo {author} {\bibfnamefont {Y.}~\bibnamefont
  {Levin}},\ }\href@noop {} {\bibfield  {journal} {\bibinfo  {journal} {J.
  Chem. Phys.}\ }\textbf {\bibinfo {volume} {136}},\ \bibinfo {pages} {194103}
  (\bibinfo {year} {2012})}\BibitemShut {NoStop}%
\bibitem [{\citenamefont {Ng}(1974)}]{Ng74}%
  \BibitemOpen
  \bibfield  {author} {\bibinfo {author} {\bibfnamefont {K.}~\bibnamefont
  {Ng}},\ }\href@noop {} {\bibfield  {journal} {\bibinfo  {journal} {J. Chem.
  Phys.}\ }\textbf {\bibinfo {volume} {61}},\ \bibinfo {pages} {2680} (\bibinfo
  {year} {1974})}\BibitemShut {NoStop}%
\bibitem [{\citenamefont {Frenkel}\ and\ \citenamefont {Smit}(2001)}]{FrSm01}%
  \BibitemOpen
  \bibfield  {author} {\bibinfo {author} {\bibfnamefont {D.}~\bibnamefont
  {Frenkel}}\ and\ \bibinfo {author} {\bibfnamefont {B.}~\bibnamefont {Smit}},\
  }\href {http://books.google.com.br/books?id=5qTzldS9ROIC} {\emph {\bibinfo
  {title} {Understanding Molecular Simulation: From Algorithms to
  Applications}}},\ Computational science series\ (\bibinfo  {publisher}
  {Elsevier Science},\ \bibinfo {year} {2001})\BibitemShut {NoStop}%
\end{thebibliography}
%

\end{document}